\documentclass[12pt]{article}

\usepackage{scicite}

\usepackage{times}

\usepackage{braket}
\usepackage{amsmath}
\usepackage{graphicx}
\usepackage[utf8]{inputenc}
\usepackage{textcomp}
\usepackage{xcolor}
\usepackage{siunitx}
\usepackage{amssymb}

\DeclareMathOperator{\arccosh}{arccosh}

\usepackage{centernot}

\newcommand{\f}[1]{\mbox{\boldmath$#1$}}
\newcommand{\fk}[1]{\mbox{\boldmath$\scriptstyle#1$}}

\newcommand{\bea}{\begin{eqnarray}}
	\newcommand{\ea}{\end{eqnarray}}
\newcommand{\eea}{\end{eqnarray}}

\newenvironment{sciabstract}{%
\begin{quote} \bf}
{\end{quote}}

\title{Particle pair creation by inflation of quantum vacuum fluctuations in an ion trap} 

\author
{M. Wittemer$^{1\ast}$, F. Hakelberg$^{1}$, P. Kiefer$^{1}$, J.-P. Schröder$^{1}$,\\
C. Fey$^{2}$, R. Schützhold$^{3,4,5}$, U. Warring$^{1}$, T. Schaetz$^{1}$\\
\\
\normalsize{$^{1}$Physikalisches Institut, Universit\"at Freiburg, Hermann-Herder-Stra{\ss}e 3, 79104 Freiburg, Germany}\\
\normalsize{$^{2}$Zentrum für Optische Quantentechnologien, Universit\"at Hamburg,}\\ \normalsize{Fachbereich Physik, Luruper Chaussee 149, 22761 Hamburg, Germany}\\
\normalsize{$^{3}$Fakult\"at für Physik, Universit\"at Duisburg-Essen, Lotharstrasse 1, 47057 Duisburg, Germany}\\
\normalsize{$^{4}$Helmholtz-Zentrum Dresden-Rossendorf, Bautzner Landstrasse 400, 01328 Dresden, Germany}\\
\normalsize{$^{5}$Institut für Theoretische Physik, Technische Universit\"at Dresden, 01062 Dresden, Germany}\\
\\
\normalsize{$^\ast$To whom correspondence should be addressed; E-mail:  matthias.wittemer@physik.uni-freiburg.de.}
}

\date{}

\begin{document} 


\baselineskip24pt


\maketitle 


%
\begin{sciabstract}
	The creation of matter and structure in our universe is currently described by an intricate interplay of quantum field theory and general relativity. 
	Signatures of this process during an early inflationary period can be observed, while direct tests remain out of reach.
	Here, we study an experimental analog of the process based on trapped atomic ions. 
	We create pairs of phonons by tearing apart quantum vacuum fluctuations. Thereby, we prepare ions in an entangled state of motion.
	Controlling timescales and the coupling to environments should permit optimizing efficiencies while keeping the effect robust via established tools in quantum information processing (QIP). 
	This might also permit to cross-fertilize between concepts in cosmology and applications of QIP, such as, quantum metrology, experimental quantum simulations and quantum computing.
\end{sciabstract}

The quantum vacuum is not empty, but filled with ubiquitous fluctuations, as implied by the Heisenberg uncertainty principle.
These fluctuations remain invisible but have observable consequences such as the Lamb shift of atomic spectral lines~\cite{Lamb1947}, van-der-Waals and Casimir forces~\cite{Casimir1948}, or the ``stimulation'' of spontaneous emission~\cite{Einstein1917,Dirac1927}.
They can be pictured as the random creation and annihilation of virtual pairs of particles and anti-particles.
Intriguingly, extreme conditions such as a rapid cosmic expansion can tear these virtual particles apart and, thereby, turn them into real particles~\cite{Schroedinger1939,Parker1969}.
These particle pairs represent lasting excitations of the quantum field and their creation generates quantum entanglement at large distances.
Similar mechanisms cause the Sauter-Schwinger effect~\cite{Sauter1931} and Hawking radiation~\cite{Hawking1974}.
In addition, pair-creation can be realized by parametric driving, e.g., via the dynamical Casimir effect\cite{Moore1970}.
According to our standard model of cosmology, the tearing apart of fluctuations by an expanding space-time during the inflationary period of the early universe explains the creation of the seeds for structure formation~\cite{Mukhanov1992}.
Realizing such extreme conditions in the laboratory remains elusive, but some analog features have been observed in several experimental platforms~\cite{Belgiorno2010,Lahav2010,Wilson2011,Weinfurtner2011,Jaskula2012,Laehteenmaeki2013,Steinhauer2016,Euve2016,Eckel2018}.
However, preserving the fragile quantum dynamics described above requires fast control of the system on the level of single quanta, as well as close to ideal isolation from the environment.
Trapped atomic ions are well suited to study fundamental quantum dynamics as they feature unique fidelities in preparation, control, and detection of quantum states~\cite{Leibfried2003,Wineland2013}.

Here, we present the creation of pairs of particles, more precisely phonons, by inflating quantum vacuum fluctuations of the motion of ions in their storage field.
This is accompanied by the creation of spatial entanglement.
We explain the process as an experimental analog to an inflationary period of an early universe~\cite{Alsing2005,Schuetzhold2007,Fey2018}.

For simplicity of the analog, let us consider a scalar (e.g., inflaton or Higgs) field $\Phi$ in 1+1 dimensions (other fields behave analogously, see~\cite{Supp}). 
In terms of the conformal time $t$, the expanding space-time during cosmic inflation can be described by the Friedmann-Lema\^{\i}tre-Robertson-Walker metric $ds^2 = a^2(t)\left[c^2dt^2-dx^2\right]$, where the time-dependent scale parameter $a(t)$ governs the cosmic expansion. 
After a spatial Fourier transform $\Phi(t,x)\to\phi_k(t)$, the equation of motion for a specific mode $k$ reads 
\begin{align}
	\label{eom}
	\ddot{\phi}_{{k}} + \left[c^2k^2 + a^2(t)\frac{m^2c^4}{\hbar^2} \right] \phi_{{k}} = 
	\ddot{\phi}_{{k}} + \Omega_{{k}}^2(t) \phi_{{k}} = 0
	\,,
\end{align}
where $c$ denotes the speed of light, $\hbar$ is the reduced Planck constant, while $k$ labels the related harmonic potential and identifies the momentum $\hbar k$ of one excitation with mass $m$.
The squared frequency $\Omega_k^2(t)$ contains the internal (propagating) contribution 
$c^2k^2$ as well as the external dynamics $\propto a^2(t)$ in the second term.
Initially ($\Omega_k\approx ck$), the internal dynamics dominates and the mode $\phi_k$ oscillates freely, staying close to its ground state.
During the course of inflation, $a(t)$ grows and the cosmic expansion stretches the physical wavelength $\lambda(t)=2\pi a(t)/k$ of the mode. 
If the external variation $a^2(t)$ is sufficiently fast to dominate the dynamics, i.e., the temporal change $\dot\Omega_k$ is not small compared to $\Omega_k^2$, the quantum state departs from the ground state, cf. Figs.~1A and B. 
The mode transforms to an excited, squeezed state, which corresponds to the creation of pairs of particles.
Thereby, the cosmic expansion tears apart the initial quantum vacuum fluctuations and turns them into pairs of particles with opposite momenta $\pm \hbar k$, generating quantum entanglement at large distances~\cite{Martin2012}, cf. Fig.~1C.
Relevant parameters for the number of particle pairs created are the scale parameter $a(t)$ and the relative rate of frequency change $\dot{\Omega}_k/\Omega_k^2$. 
The above mechanism is universal and applies to any quantum harmonic oscillator subjected to large and rapid change of its frequency $\omega_a(t)$, which causes squeezing of its wavefunction $\psi_a$.
In an ion trap, the motional (phonon) modes represent the $\phi_k$, while changes of the trapping potential mimic the cosmic expansion $a(t)$ and, thus, stimulate the creation of pairs of phonons~\cite{Supp}.

\begin{figure}[hb]\centering
	\includegraphics{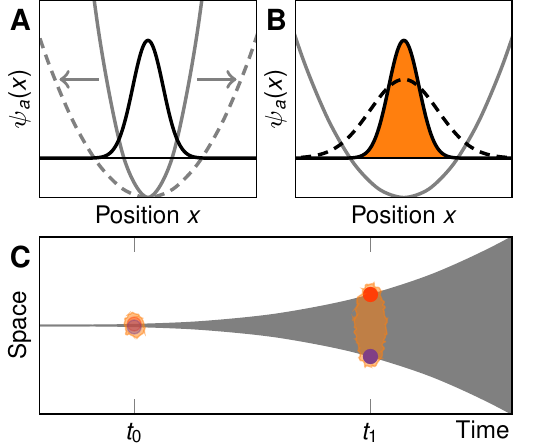}
	\caption{
		\textbf{Particle creation out of quantum fluctuations in harmonic oscillators and the early universe.}
		(\textbf{A})
		The ground state wavefunction $\psi_a(x)$ (black solid line) of a harmonic oscillator has finite energy and spatial spread (variance) $\left(\Delta x\right)^2 \propto 1/\omega_a$. 
		When $\omega_a$ is changed to a different value (dashed parabola), $\psi_a(x)$ evolves accordingly.
		(\textbf{B}) 
		If the change of $\omega_a$ is fast, $\psi_a(x)$ cannot follow adiabatically and represents an excited state (solid line) of the new potential.
		It is characterized by a squeezed shape, with decreased spread $\left(\Delta x\right)^2$ compared to the new ground state (dashed line).
		This squeezing excitation corresponds to the creation of pairs of particles.
		(\textbf{C})
		Schematic of curved space-time during cosmic inflation. 
		Quantum fluctuations are depicted as a pair of virtual particles at some time $t_0$. 
		Due to inflation the two virtual particles are torn apart until their distance (related to the physical wavelength of the corresponding mode, shaded area) becomes too large for them to recombine and annihilate (at $t_1$). 
		Thereafter the two virtual particles have become real and move into opposite directions.
		However, they remain linked via quantum entanglement.
		\label{fig1}}
\end{figure}

\clearpage
In our experiments we study the motional modes of two atomic magnesium ions confined in a linear Paul trap, details on the experimental setup are given in \cite{Clos2016,Supp}.
The two ions align along the axial direction of the trap and in the following, we focus on their harmonic motion along the weaker of the radial directions.
We distinguish an in-phase (center-of-mass) mode from an out-of-phase (rocking) mode with frequency $\omega_a$.
The latter is intrinsically robust against homogeneous noise fields, and thus, well suited to implement the dynamics according to Eq.~(1).
Our apparatus allows for real-time control of a trapping voltage $U$ that tunes $\omega_a$, as depicted in Fig.~2A.
We define an analog scale parameter, postponing the internal dynamics of Eq.~(1), according to $a_a(t)=\omega_a(t)/\omega_a(0)$, see~\cite{Supp}.
In our experiments we ramp $\omega_a/(2\pi)$, down and up, spanning $\SI{2.50(1)}{\mega \Hz}$ to $\SI{0.50(1)}{\mega \Hz}$, where each ramp corresponds to an inflation of space by about 1.6 $e$-foldings, see Fig.~2B.
We ramp $U$ within $t_\text{ramp}=\SI{1}{\micro \second}$, yielding a non-adiabatic evolution of $\psi_a$, characterized by $\dot{\omega}_a/\omega_a^2 \approx 5$, cf. Fig.~2C.
Numerical simulations~\cite{Supp} illustrate the generation of squeezing with an amplitude $r$ during a single ramp of $\omega_a$, see Fig.~2D, which corresponds to an instantaneous average particle number $\bar{n}_\text{sq}=\sinh^2\left(r\right)$.
Due to the non-instantaneous switching of $\omega_a$, the Wentzel-Kramers-Brillouin (WKB) phase $\varphi_a(t) = \int_{0}^{t}\omega_a(t')dt'$ of $\psi_a$ evolves significantly during $t_\text{ramp}$ and leads to an oscillatory accumulation of $r$ and $\bar{n}_\text{sq}$.

\begin{figure}[hb]\centering
	\includegraphics{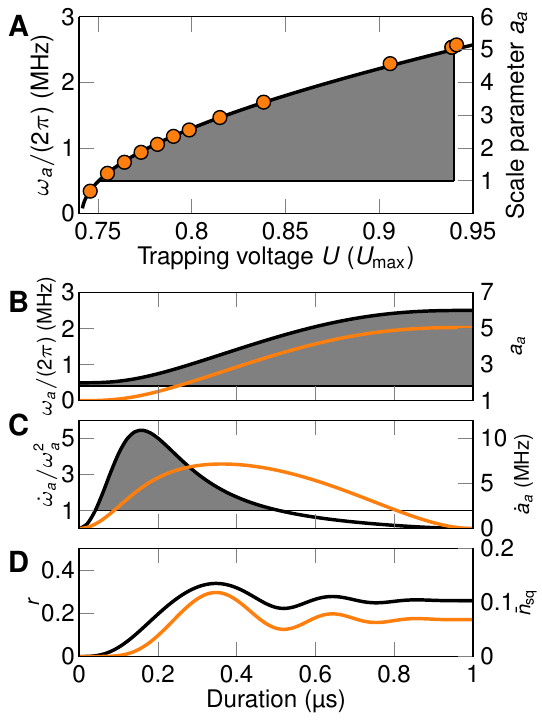}
	\caption{
		\textbf{Accessible experimental parameter regime in our setup.}
		(\textbf{A})~ 
		Frequency of the harmonic oscillator mode $\omega_a$ as a function of $U$, data points depict experimental results (error bars smaller than symbol size), the solid line shows a model fit to the data.
		The selected frequency range (shaded area) is compared to the analog scale parameter $a_a$ (right axis), which corresponds to 1.6 $e$-foldings of space.
		(\textbf{B})~
		Real-time switching of $\omega_a$ (black line) within $t_\text{ramp}=\SI{1}{\micro\second}$ and corresponding time-evolution of $a_a$ (orange line, right axis).
		(\textbf{C})~ 
		The fast variation $\dot{\omega}_a/\omega_a^2$ (black line) is accompanied by significant time derivatives of the scale parameter $\dot{a}_a$ (orange line, right axis).
		(\textbf{D})~ 
		Numerical simulation of the quantum dynamics.
		Quantum fluctuations are inflated and squeezing $r$ (black line) emerges.
		This corresponds to the pairwise creation of $\bar{n}_\text{sq}$ phonons (orange line, right axis).
		\label{fig2}}
\end{figure}

\clearpage
In a first experimental realization (protocol depicted in Fig.~3A) we cool all modes close to their motional ground state, for the rocking mode with a measured, residual thermal $\bar{n}_\text{th}=0.03(6)$, see Fig.~3B.
In order to increase the number of created particle pairs, we apply a sequence of two ramps to form one pulse: After ramping down $\omega_a$ and an appropriate duration $t_\text{hold}$, $\omega_a$ is ramped back up.
Finally, we readout the motional state by mapping it onto the electronic state of one ion~\cite{Leibfried2003,Supp}, and reconstruct the individual phonon number distribution $P_n$ (data points in Figs.~3B and C).
Typically, experimental sequences are repeated $40000$ times for a given set of parameters.
In comparison to the initial state, the data show increased populations for higher phonon number states $n$, indicating mode excitation due to the pulse.
To evaluate our results, we decompose the excitation into contributions by $r$, a coherent displacement $\alpha$, and a thermal share $\bar{n}_\text{th}$. 
We fit the data with a parametrized phonon number distribution $P_n^\text{par}(r,\alpha,\bar{n}_\text{th})$ of a Gaussian state (bar chart in Fig.~3C), with $r=0.54(8)$ and $|\alpha|=0.88(6)$.
Since all heating effects remain negligible within experimental sequences we keep $\bar{n}_\text{th}$ at the initial value 0.03.
We attribute the coherent displacement to residual static differential stray fields in our setup~\cite{Supp}.
Both excitation amplitudes, $r$ and $|\alpha|$, depend on the phase $\varphi_a$ accumulated during the pulse, i.e., they are determined by $t_\text{ramp}$, $t_\text{hold}$, and $\omega_a(t)$, see~\cite{Supp}.

\begin{figure}[hb]\centering
	\includegraphics{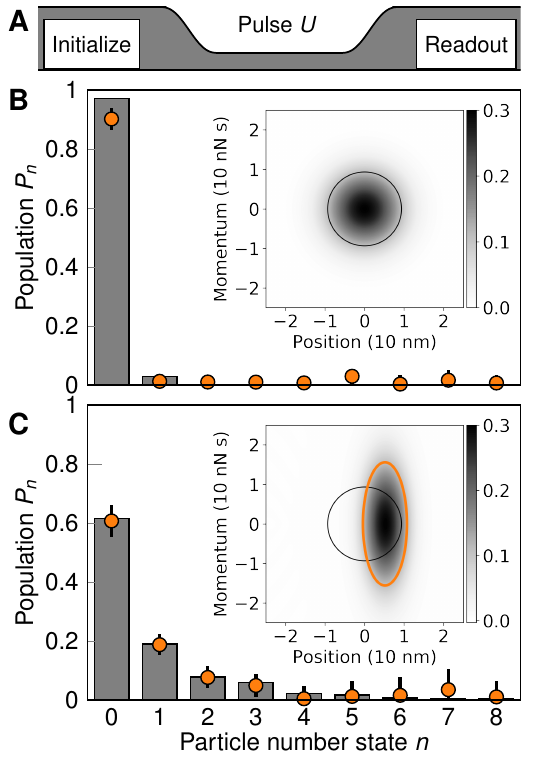}
	\caption{
		\textbf{Experimental results for particle creation accumulated during one pulse.}
		(\textbf{A})~ 
		Schematic of the experimental pulse sequence (not to scale).
		(\textbf{B})~ 
		Initial state of $\psi_a$, close to the ground state. 
		Data points are the result of 40000 realizations and indicate the individual populations of the number states, error bars depict, for illustration purposes, three standard errors of the mean (sem). 
		The inset illustrates the vacuum state in its phase space representation (position and momentum) by the corresponding Wigner function~\cite{Leibfried2003}.
		It features equal variances $\left(\Delta x\right)^2$ and $\left(\Delta p\right)^2$ emphasized by the circle depicting its 1/$e$-amplitude, where $\left(\Delta x\right)^2 \cdot \left(\Delta p\right)^2 = \hbar^2/4$ reflects the minimum uncertainty set by Heisenberg’s principle.
		(\textbf{C})~
		The final motional state shows significant excitations.
		From the model fit to our data, we reconstruct the Wigner function (inset, phases omitted for clarity) that deviates from the initial vacuum state (circle).
		It is coherently displaced from the origin by $|\alpha|$, to an offset of $\approx\SI{5}{\nano\meter}$, and squeezed by $\approx\SI{4.7}{\decibel}$ on one variance in expense of the other (ellipse for 1/$e$-amplitude), still remaining close to the Heisenberg limit.
		\label{fig3}}
\end{figure}

\clearpage
In order to reverse the coherent displacement, we perform purifying echo sequences consisting of two pulses, separated by a duration of free evolution $t_\text{free}$, see Fig.~4A.
To illustrate the echo effect of the second pulse we depict the corresponding phase space animation in Fig.~4B.
The initial thermal state (i) is squeezed and coherently displaced by the first pulse (ii), similar to Fig.~3C. 
During $t_\text{free}$ the state oscillates, i.e., rotates on a circular path in phase space (iii). 
$\psi_a$ accumulates two distinct phases corresponding to excitations $r$ and $\alpha$ with frequencies $2\omega_a(t_\text{free})$ and $\omega_a(t_\text{free})$, respectively.
Ideally, the second pulse is applied after a duration $\pi/\omega_a(t_\text{free})$ (or odd multiples), where the coherent displacement has evolved from $\alpha$ to $-\alpha$ (iv), and the final state after the second pulse (v) is characterized by $\alpha\approx0$ and enhanced squeezing~\cite{Supp}.
We experimentally realize the two-pulse sequence for variable $t_\text{free}$, perform the motional state analysis, and depict resulting squeezing and coherent excitations in Fig.~4C.
As indicated by the data, for $t_\text{free} \approx \SI{30.2}{\micro \second}$, the squeezing is significantly increased, while the coherent excitation is reduced with respect to Fig.~3C.
Accordingly, in Fig.~4D we depict the $P_n$ for $t_\text{free} = \SI{30.2}{\micro \second}$.
Here the squeezing is directly evidenced by increased populations $P_n$ for even states only, while odd states remain nearly unpopulated.
This corresponds to the creation of pairs of phonons.
For example, we detect single phonon pairs ($n=2$) in $P_2\approx\SI{20}{\percent}$ of our realizations ($\approx8000$ in total), and the Wigner function indicates a suppression of the variance $\left(\Delta x\right)^2$ by $\approx\SI{7.2}{dB}$.

Moreover, features of entanglement of the particle pairs can be witnessed in the spatial degrees of freedom of our two-ion crystal~\cite{Fey2018}.
To first order, the wavefunction's squeezing excitation can be visualized in a simplifying way with a term $\propto r \left(\ket{{}^\bullet\,{}_\bullet}+\ket{{}_\bullet\,{}^\bullet}\right)$, resembling a Schrödinger cat state.
Here, $\ket{{}^\bullet\,{}_\bullet}$ and $\ket{{}_\bullet\,{}^\bullet}$ indicate the ions' non-classical anti-correlation, that can be identified with particles of momenta $+\hbar k_a$ and $-\hbar k_a$, respectively. 
We note that the creation of multiple particle pairs per state ($n=4,6,8$) that we observe evidences the bosonic character of the underlying quantum statistics of phonons~\cite{Supp}.
By employing the criterion in \cite{Serafini2004} for Gaussian states we quantify the entanglement of formation $E_F\approx0.41$, significantly increased with respect to the intrinsic vacuum state entanglement $E_F\approx10^{-5}$, see~\cite{Supp}.
In future studies, we may be able to effectively decouple both ions, while preserving their non-classically correlation, and individually map the entanglement onto the ions' electronic degree of freedom~\cite{Retzker2005,Jost2009}.

In general, many fundamental predictions rely on the premise that the laws of quantum theory remain valid, not only, across very different length and energy scales, but also during drastic changes of the external conditions.
Here, we tested this premise in our trapped ion analog.
We can continue these studies for generalized analogies involving scalar or vector fields in different dimensions.
Depending on the interpretation of the time coordinate (e.g., transformation from proper to conformal time), we may simulate different dynamics of the related scale parameter~\cite{Supp} and investigate its consequences.
For this, we can further tune durations and shapes of the ramps of our potential or add parametric driving, e.g., to simulate inflation followed by a cosmological (p)re-heating phase corresponding to an oscillatory particle creation.
Our platform may allow studying the causal connections of squeezing, pair creation, and entanglement, e.g., in the context of Hawking radiation due to surface gravity, the crossing of cosmic horizons during inflation, and the Sauter-Schwinger effect relating to high-field lasers~\cite{Supp}.
Considering the controlled coupling of our closed quantum system to environments and noise fields might permit simulating realistically extended analogs.
Further, our method represents a novel tool to create squeezing that can assist in gaining sensitivity for quantum metrology applications, cf.~\cite{Burd2018}.
In addition, squeezing has also been proposed to substantially enhance effective spin-spin interactions, required for experimental simulations of quantum spin models~\cite{Cirac2012,Ge2019}, and has recently been used to implement qubits for QIP in the motional states of trapped ions~\cite{Fluehmann2019}.
Finally, our results regarding squeezing and purification from coherent excitations should be considered when implementing (multi-ion) entangling gates on multiplex trap architectures, where rapid changes of potential landscapes are required for scaling towards a universal quantum computer~\cite{Cirac2000,Kielpinski2002}. 

\begin{figure}[hb]\centering
	\includegraphics{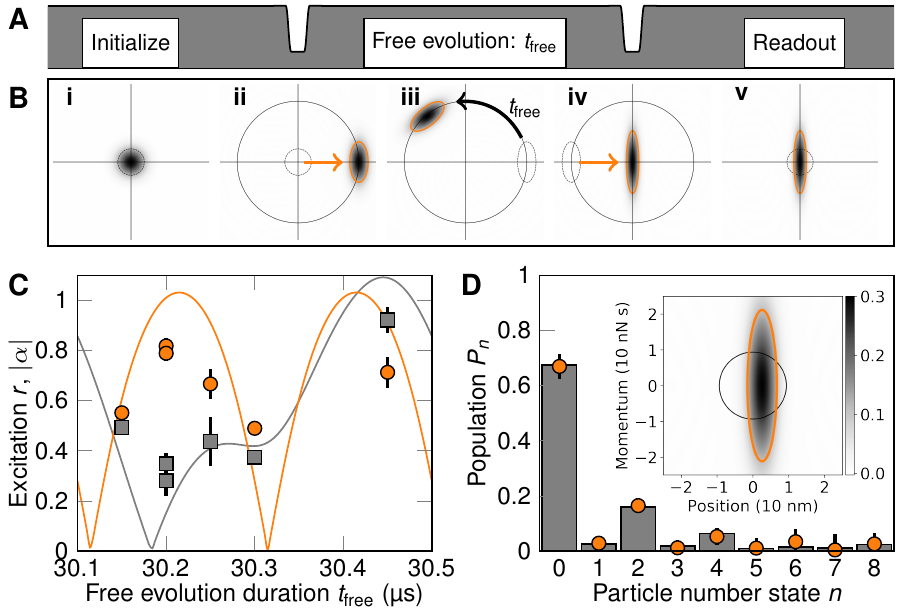}
	\caption{
		\textbf{Purifying pair creation with an echo sequence.}
		(\textbf{A})~ 
		Experimental sequence (not to scale) consisting of two pulses separated by a free evolution duration $t_\text{free}$.
		(\textbf{B})~ 
		Phase space animation illustrating the anticipated echo effect:
		(i) The initial vacuum state, (ii) squeezed and displaced by the first pulse, (iii) rotates on a circular path in phase space.
		(iv) For an optimal duration $t_\pi=\pi/\omega_a$ (or odd multiples) the residual coherent displacement is reversed by the second pulse.
		(v) The final state shows enhanced squeezing excitation.
		(\textbf{C})~
		Experimental results for squeezed (circles) and coherent (squares) excitation as a function of $t_\text{free}$, error bars indicate the sem. 
		Numerical simulations (lines) indicate the different oscillation rates of $r(t_\text{free})$ and $\alpha(t_\text{free})$, see~\cite{Supp}.
		(\textbf{D})~
		Phonon number distribution (data points, error bars indicate three sem) for $t_\text{free} = \SI{30.2}{\micro\second}$. 
		From a model fit (bars) we extract $r=0.83(8)$ and $|\alpha|=0.29(15)$ and the corresponding Wigner function (inset, phases omitted for clarity).
		The squeezed state evidences entanglement in the spatial degrees of freedom of the two ions, represented by the superposition $\ket{{}^\bullet\,{}_\bullet}+\ket{{}_\bullet\,{}^\bullet}$, see main text.
		\label{fig4}}
\end{figure}

\clearpage

\section*{Acknowledgments}
This work was supported by the Deutsche Forschungsgemeinschaft (DFG) [237456450 and 278162697 (SFB 1242)]. 




\clearpage


\section*{Supplementary Materials}

\renewcommand{\thefigure}{S\arabic{figure}}
\setcounter{figure}{0}
\setcounter{equation}{0}
\renewcommand{\theequation}{S\arabic{equation}}

\noindent Materials and Methods\\
\noindent Supplementary Text\\
\noindent Figures S1 to S4\\
\noindent References \textit{(37-47)}\\

\section{Analogies}

In this Section, we provide a brief (and incomplete) list of scenarios for particle pair creation by tearing apart quantum vacuum fluctuations and show how they can be mapped onto harmonic oscillators with time-dependent frequencies which undergo large and rapid changes. 
We focus here on emphasizing analogies that might help to understand causal connections, similarities, and differences of the effect of particle creation in different contexts. 
For a detailed description, please refer to \cite{Birrell1982}. 
At the end of this Section we provide a summary in a dictionary, translating the physics and its corresponding terminology into the language of trapped ions.

\subsection*{Proper Time}

In 3+1 dimensions, the expanding space-time during cosmic inflation can be described by the Friedmann-Lema\^{\i}tre-Robertson-Walker metric 
\bea
\label{FLRW-proper}
ds^2=c^2d\tau^2-a^2(\tau)d\f{r}^2
\ea
in terms of the proper time $\tau$, where the time-dependent scale parameter $a(\tau)$ governs the cosmic expansion. 

After a spatial Fourier transform $\Phi(\tau,\f{r})\to\Phi_{\fk{k}}(\tau)$ and a field redefinition $\phi_{\fk{k}}(\tau)=\Phi_{\fk{k}}(\tau)\sqrt{a^3(\tau)}$, the equation of motion $(\Box+m^2c^2/\hbar^2)\Phi=0$ of a scalar field simplifies to 
\bea
\label{oscillator-proper}
\ddot\phi_{\fk{k}}
+
\left[
\frac{m^2c^4}{\hbar^2}
+
\frac{c^2\f{k}^2}{a^2}
-
\frac32\left(\frac{\ddot a}{a}+\frac12\frac{\dot a^2}{a^2}\right) 
\right]
\phi_{\fk{k}}=0
\,.
\ea
In this representation, one can directly read off the stretching of physical wavelengths $\lambda=2\pi a/|\f{k}|$ due to the cosmic expansion in the term $c^2\f{k}^2/a^2$. 

The Hubble rate $H$ measuring the rapidity of the cosmic expansion is given by $H=\dot a/a$. 
For example, the de~Sitter space-time is described by a constant Hubble rate $H$ and corresponds to an exponential growth of the scale parameter $a(\tau)\propto \exp(H\tau)$.
In this case, the last term in the square brackets in~\eqref{oscillator-proper} is simply given by $9H^2/4$. 
Its competition with the contribution $c^2\f{k}^2/a^2$ from the internal (propagating) dynamics marks the event of horizon crossing:
For wavelengths $\lambda=2\pi a/|\f{k}|$ much smaller than the Hubble radius $c/H$ (which determines the size of the cosmic horizon), the modes oscillate nearly freely. 
However, at some point in time the wavelength has been stretched enough by the cosmic expansion such that it exceeds the Hubble radius. 
Afterwards, the modes cannot oscillate anymore due to loss of causality and thus they cannot adapt to the changing external conditions, i.e., adiabaticity breaks down. 

\subsection*{Conformal Time}

In terms of the conformal time coordinate $t$ used in the manuscript, the 3+1 dimensional Friedmann-Lema\^{\i}tre-Robertson-Walker metric reads 
\bea
\label{FLRW-conformal}
ds^2=a^2(t)\left[c^2dt^2-d\f{r}^2\right]
\,.
\ea
Again, a field redefinition $\phi(t,\f{r})=a(t)\Phi(t,\f{r})$ and a spatial Fourier transform yield  
\bea
\label{oscillator-conformal}
\ddot\phi_{\fk{k}}
+
\left[
a^2\,\frac{m^2c^4}{\hbar^2}
+
c^2\f{k}^2
-
\frac{\ddot a}{a}
\right]
\phi_{\fk{k}}=0
\,.
\ea
Both Equations~\eqref{oscillator-proper} and \eqref{oscillator-conformal} are analog to harmonic oscillators with time-dependent frequencies, but the explicit expressions vary due to the different time coordinates. 
Note that also the explicit functional form of the time-dependence of the scale parameter $a$ changes with such a coordinate transformation. 
For example, the de~Sitter space-time corresponds to $a(\tau)\propto\exp(H\tau)$ in proper time $\tau$, 
but to $a(t)\propto1/t$ in conformal time $t$.
Thus, the space-time simulated by the trapped ions depends on which time coordinate is identified with the laboratory time. 

\subsection*{Vector Field}

In contrast to the scalar (spin-zero) field $\Phi$ above, we may also consider a vector (spin-one) field $A^\mu$ such as a Proca field as described by the Proca equation of motion $(\nabla_\mu\nabla^\mu+m^2c^2/\hbar^2)A^\nu=0$, where we have used the generalized Lorenz gauge $\nabla_\mu A^\mu=0$. 
After a mode decomposition $A^\mu(t,\f{r})\to A_{\fk{k},\sigma}(t)$, where $\sigma$ labels the polarization, we arrive at (in terms of the conformal time)
\bea
\label{vector-field}
\ddot A_{\fk{k},\sigma}+
\left[
a^2\,\frac{m^2c^4}{\hbar^2}
+
c^2\f{k}^2
\right]
A_{\fk{k},\sigma}=0
\,.
\ea
We see that the time-dependent scale factor $a(t)$ only enters via the mass term.
As a result, photons are not created (as they are massless), which is a direct consequence of the conformal invariance of the electromagnetic field in 3+1 dimensions.
However, one could create massive vector bosons such as $W^\pm$ or $Z^0$.  
This would require sufficiently large frequency scales in $a(t)$ which reach or even exceed the mass scale set by $m_W$ or $m_Z$.
These large frequency scales in $a(t)$ arise naturally in the early universe, but one should check whether the above description in Eq.~\eqref{vector-field} is a valid approximation in this period (symmetry breaking etc.). 
On the other hand, for Dirac (e.g., spin-1/2) fields $\Psi$, a quite analog mapping is possible.  
However, these are fermionic fields, while the vibrational excitations of the ions are of bosonic nature, i.e., the quantum statistics is then not directly captured. 

\subsection*{Sauter-Schwinger Effect}

As the next example, let us consider the Sauter-Schwinger effect, i.e., the creation of charged particle pairs by a strong electric field $\f{E}$. 
We assume a spatially homogeneous but time-dependent electric field $\f{E}(t)$ which can be represented by the vector potential $\f{A}(t)$ in temporal gauge via $\f{E}(t)=\f{\dot A}(t)$. 
The equation of motion for a complex scalar field $\Phi$ reads (in flat space-time)
\bea
\left[
\partial_t^2-c^2\left(\f{\nabla}-i\frac{q}{\hbar}\f{A}\right)^2+\frac{m^2c^4}{\hbar^2}
\right]
\Phi=0
\,,
\ea
where $q$ denotes the charge of the scalar field. 
A spatial Fourier transform simplifies this equation to 
\bea
\label{KFG-k}
\ddot\phi_{\fk{k}}+
\left[c^2\left(\f{k}-\frac{q}{\hbar}\f{A}\right)^2+\frac{m^2c^4}{\hbar^2}\right]\phi_{\fk{k}}=0
\,.
\ea
Again we find the analogy to a harmonic oscillator with a time-dependent frequency. 

In order to observe the non-perturbative Sauter-Schwinger effect associated to the strong-field regime, the electric field should be strong enough and slowly varying. 
This requirement can be quantified by means of the dimensionless (relativistic) Keldysh parameter 
\bea
\label{Keldysh}
\gamma=\frac{\omega mc}{qE}\sim\frac{mc}{qA}  
\,,
\ea
where $\omega$ is a characteristic frequency scale of the electric field. 
Small Keldysh parameters $\gamma\ll1$ correspond to the  non-perturbative (strong-field) tunneling regime of the Sauter-Schwinger effect while large Keldysh parameters $\gamma\gg1$ signify the perturbative 
(weak-field) multi-photon regime. 
Comparing Eqs.~\eqref{KFG-k} and \eqref{Keldysh}, we see that reaching the non-perturbative strong-field regime $\gamma\ll1$ of the Sauter-Schwinger effect requires large variations of the harmonic oscillator frequency in Eq.~\eqref{KFG-k}.
In contrast, small changes of the harmonic oscillator frequency in Eq.~\eqref{KFG-k} such as in parametric down conversion in quantum optics (see also \cite{Burd2018}) correspond to the perturbative (weak-field) regime $\gamma\gg1$. 

\subsection*{Hawking Radiation}

As our final example, let us consider Hawking radiation, i.e., black-hole evaporation. 
In terms of the Kruskal light-cone coordinate $U$, the black-hole metric in the (1+1 dimensional) near-horizon region reads 
\bea
ds^2=2ce^{\kappa t}dt\,dU-e^{2\kappa t}dU^2
\,,
\ea
where $\kappa$ is the surface gravity of the black hole and determines the Hawking temperature via $T_{\rm H}=\hbar\kappa/(2\pi k_{\rm B})$, see, e.g., \cite{Schutzhold2013}.  
The term $e^{\kappa t}$ describes the tearing apart of waves by the strong tidal forces near the horizon -- which is quite analog to particle creation by an expanding universe. 
Introducing the new time coordinate $T$ (not to be confused with the Hawking temperature $T_{\rm H}$)
similar to the conformal time via $e^{-\kappa t}=-\kappa T$, the above metric reads 
\bea
ds^2=\frac{2cdT\,dU-dU^2}{(\kappa T)^2}
\,,
\ea
where the global pre-factor $1/(\kappa T)^2$ now plays the role of the scale parameter $a^2(T)$.

The equation of motion for a scalar field $\Phi$ reads 
\bea
\left[\partial_T^2+2c\partial_T\partial_U+a^2(T)\,\frac{m^2c^4}{\hbar^2}
\right]\Phi=0
\,.
\ea
Finally, a Fourier expansion into modes $e^{iKU}$ and a field redefinition $\varphi_K(T)=e^{icKT}\phi_K(T)$ yields 
\bea
\left[\partial_T^2+c^2K^2+a^2(T)\,\frac{m^2c^4}{\hbar^2}
\right]\varphi_K(T)=0
\,,
\ea
which is formally the same equation as in the manuscript. 
Thus, adjusting the time-dependence of the mode frequency according to $\omega^2_K(T)=c^2K^2+m^2c^4/(\hbar\kappa T)^2$ would mimic the tearing apart of quantum vacuum fluctuations 
near the black hole horizon by the strong gravitational field (as characterized by the surface gravity $\kappa$). 

However, one should bear in mind that a full treatment of particle creation -- especially in this scenario -- must also include a consideration of the asymptotic regions (not just the vicinity of the horizon). 
Nevertheless, the tearing apart of quantum vacuum fluctuations is the most vital ingredient for Hawking radiation, see, e.g.,~\cite{Schutzhold2013}.  
The entanglement generated in this process explains the thermal nature of the outgoing radiation:
If we consider only one partner of the pair forming an entangled state, its density matrix corresponds to a mixed (thermal) state. 

\subsection*{Dictionary}

Let us discuss the analogy to the ion trap in more detail.
Linearizing the radial displacements $\delta q_i$ of the ions (labeled by $i$), we obtain the equations of motion 
\bea
\delta\ddot{q}_i+\omega^2_{\rm rad}(t)\delta q_i+\sum\limits_j M_{ij}\delta q_j=0
\,,
\ea
where $\omega^2_{\rm rad}(t)$ encodes the radial confinement controlled by the time-dependent trap potential and the matrix $M_{ij}$ describes the Coulomb interactions between the ions. 
Diagonalizing this matrix with the eigenvalues $\lambda_I$ (not to be confused with the wavelength) labeled 
by the index $I$ is then analog to the Fourier transform above and we obtain the mode equations 
\bea
\label{ion-modes}
\delta\ddot{q}_I+\omega^2_{\rm rad}(t)\delta q_I+\lambda_I\delta q_I=0
\,,
\ea
which have a similar form as in the subsections above. 
The Coulomb interaction strength $\lambda_I$ is then analog to the internal (propagating) dynamics $c^2\f{k}^2$.
In both cases, they are responsible for generating entanglement.
The trap potential $\omega^2_{\rm rad}$ simulates the mass term $m^2c^4/\hbar^2$ plus possible potential terms $\propto \dot a^2$ or $\propto\ddot a$  as in Eq.~\eqref{oscillator-proper}. 
Its time-dependence $\omega^2_{\rm rad}(t)$ represents the external influence and is then analog to the cosmic expansion.
As shown above, the radial displacements $\delta q_i$ of the individual ions correspond to the field $\Phi$ (or $A^\mu$ etc.) in position representation while the modes $\delta q_I$ (such as the center-of-mass or rocking mode) are analog to the Fourier modes $\phi_{\fk{k}}$.

Holding the axial positions of the ions fixed, the Coulomb interaction strength $\lambda_I$ stays constant. 
Thus, when identifying $\lambda_I$ with the internal dynamics $c^2\f{k}^2$, Eq.~\eqref{ion-modes} is most similar to the evolution equation~\eqref{vector-field} of a vector field in 3+1 dimensions, or Eq.~(1) of a scalar field in 1+1 dimensions, where the additional potential terms $\propto \dot a^2$ or $\propto\ddot a$ are absent. 
This again shows that the explicit details of the equations of motion -- albeit they are all qualitatively similar -- depend on the field (e.g., $\Phi$ or $A^\mu$), the number of dimensions and the used time coordinate etc.
For the vector field $A^\mu$, we could identify the two transversal polarizations $\sigma$ with the two radial directions. 

Identifying the harmonic oscillator potential $\omega^2(t)$ with $c^2\f{k}^2+a^2(t)m^2c^4/\hbar^2$ as in the manuscript and in Eq.~\eqref{vector-field}, we see that a relative change of $\omega(t)$, such as a certain number of $e$-foldings, implies an even larger relative change of $a(t)$, i.e., more $e$-foldings, because the remaining constant term $c^2\f{k}^2$ is positive. 
Thus, the reported 1.6 $e$-foldings of $\omega(t)$ imply, in principle, more than 1.6 $e$-foldings of space $a(t)$.

Pairs of phonons (as vibrational excitation quanta) of the ions represent pairs of particles as lasting excitations of the field $\Phi$ (or $A^\mu$ etc.). 
Finally, the entanglement of distant space points created during inflation corresponds to the motional entanglement of the two ions.  
After separating the ions and reading out the motional states individually, the entanglement would manifest itself as a thermal (i.e., mixed) state with a temperature set by the squeezing parameter -- in complete analogy to Hawking radiation which appears thermal because one can only observe the one partner of the particle pair escaping to infinity (the other one is falling into the black hole). 

\section{Experimental methods}

Our experimental setup features a linear radio-frequency (rf) ion trap and most of the relevant techniques employed here are described in detail in~\cite{Clos2016,Leibfried2003}.
We trap and manipulate different isotopes of Mg$^+$ in a common electric potential. 
Radial confinement is dominated by an rf potential oscillating at $\Omega_\text{rf}/(2\pi) \approx \SI{56}{\mega\hertz}$, while fine tuning along the radial directions is accomplished by dedicated control voltages applied to six control electrodes.
In addition, we use these electrodes to adjust the axial confinement, to which rf fields contribute effectively less than a few percent.
Our rf generator setup driving the two rf electrodes includes a signal generator with fixed amplitude, an rf mixer controlled via an arbitrary waveform generator~\cite{Bowler2013} (with output amplitude $U_\text{c}$), a high power amplifier (output power $\approx 40$\,dBm), a bandpass filter (\SI{800}{\kilo\hertz} bandwidth) and a helical resonator with a loaded quality factor $Q \approx 100$.
With our real-time data acquisition system, with a \SI{10}{\nano\second} timing resolution, we tune and ramp the applied rf voltage amplitude $U$ between \SI{0}{\volt} and \SI{800}{\volt}.

We use a single ${}^{25}\text{Mg}^+$ in our trap to calibrate and optimize control parameter settings, i.e., tuning of offset electric fields and curvatures, while for the main sequences (main text), we add one ${}^{26}\text{Mg}^+$ and form a strongly coupled hybrid crystal.
We choose the ${}^{25}\text{Mg}^+$ to implement a pseudo spin system (internal/electronic degree of freedom) on two suitable S$_{1/2}$ hyperfine ground states, labeled $\ket{\downarrow}$ and $\ket{\uparrow}$.
We employ Doppler cooling and subsequent resolved sideband cooling via two-photon stimulated-Raman transitions~\cite{Leibfried2003,Schmitz2009,Clos2016,Wittemer2017} applied to the ${}^{25}\text{Mg}^+$.
In this way, we cool up to six motional modes (external degrees of freedom) and yield low mean thermal occupation numbers.
We employ established techniques to map motional state distributions to the spin states, and electron shelving to distinguish $\ket{\downarrow}$ (bright state) from  $\ket{\uparrow}$ (dark state), cf.~\cite{Leibfried2003}.
The mapping technique is based on the coherent coupling of the internal and external degrees of freedom via resonant transitions $\ket{\downarrow,n}\leftrightarrow\ket{\uparrow,n+1}$ (blue sideband) and $\ket{\uparrow,n}\leftrightarrow\ket{\downarrow,n-1}$ (red sideband), where $\ket{n}$ denotes the corresponding number state of the individually addressed mode.

In a first series of calibration measurements, we record the variation of motional frequencies as a function of $U$, results are shown in Fig.~S1A.
We distinguish two radial modes, the high-frequency (hf) mode $\omega_\text{hf}$ and the mid-frequency (mf) mode $\omega_\text{mf}$ (blue squares), from the axial (low-frequency, lf) mode $\omega_\text{lf}$ (green diamonds), solid lines depict model fits to the data.
In addition, we estimate the frequency variation of the rocking mode $\omega_a\approx\sqrt{\omega_\text{mf}^2 - \omega_\text{lf}^2}$ (orange circles).
During the course of our experimental runs, we monitor changes of $U$ with a dedicated pickup electrode that is capacitively coupled to the rf electrodes, and the corresponding pickup signal amplitude is denoted by $U_p$. 
Typically, we program a sixth-order smooth-step waveform $U_c(t)$ to modulate $U$, with fixed ramp duration $t_\text{ramp}=\SI{1}{\micro\second}$ (cf. Fig.~2B), pulsing between high amplitude $U_\text{high}$ and low amplitude $U_\text{low}$.
We record $U_p(t)$ and the input waveform $U_c(t)$ to characterize the real-time switching of $U$.
In Figure~S1B, we depict corresponding results of a single pulse (two ramps), taking into account the calibration measurements in Fig.~S1A.
Due to the finite bandwidth of our rf generator setup (including all wiring), the waveform $U_p$ is shifted and distorted with respect to $U_c$.
We account for the shift by an appropriate delay $t_\text{delay}$ in our experimental sequences. 
While the distortion can affect the (quantitative) details of the mode dynamics, general (qualitative) features are robust and final squeezing amplitudes vary by a few percent only, see below.

In a second series of calibration runs, we benchmark the real-time switching capabilities of our setup with a spin-motional echo sequence~\cite{Leibfried2002} on the mf mode with an interferometric duration (arm length) $t_\text{se} = 2\,t_\text{ramp} + t_\text{hold} + t_\text{delay}$, see Fig.~S1C. 
We initialize the state $\ket{\downarrow,n_\text{mf}=0}$ and, subsequently, create a coherent superposition $\ket{\downarrow,0}+\ket{\uparrow,1}$ with a $\pi/2$-pulse on the blue sideband.
During the first arm, we ramp down $U$, hold at $U_\text{low}$ for a duration $t_\text{hold}$, and ramp back up to $U_\text{high}$.
Note, during $t_\text{se}$ the spin-motional superposition state accumulates a phase that is proportional to $\int_0^{t_\text{se}}\omega_\text{mf}(t)dt$.
Following an echo pulse, i.e., a $\pi$-pulse on the blue sideband, $U$ is kept high (no pulse) during the second arm and the sequence concludes with another $\pi/2$-pulse on the blue sideband, and spin-state detection.
Overall, this sequence is sensitive to differential phase accumulations of both arms, i.e., mode frequency differences can be detected. 
Repeating measurements for variable $t_\text{hold}$ and $t_\text{se}$, accordingly, we record final $P_{|\downarrow\rangle}$ and show corresponding results in Fig.~S1D. 
From a sinusoidal model fit to the data we extract a mode frequency difference of $2\pi\times\SI{1.53(3)}{\mega\hertz}$, in agreement with the static calibration values $\omega_\text{mf}(U_\text{high})-\omega_\text{mf}(U_\text{low})= 2\pi\times\SI{1.48(1)}{\mega\hertz}$, cf. Fig.~S1A.

In a third series of calibration measurements, we estimate motional heating rates~\cite{Leibfried2003} for the two-ion configuration.
In particular, we probe the mid-frequency mode and the accompanying rocking mode for $U_\text{high}$.
We record rates of $\lesssim\SI{2000}{quanta/\second}$ for the center-of-mass and $\lesssim\SI{20}{quanta/\second}$ for the rocking mode, while heating rates for the other four modes of the crystal are comparable or smaller.
For total sequence lengths of smaller than \SI{200}{\micro\second} in our main experiments, we can neglect incoherent (thermal) mode excitations on the rocking mode. 
Note, even when conservatively estimating a $60$-fold increased rate for $U_\text{low}$, the rocking mode is heated less than $\approx 0.02$ quanta during our sequences.
This reflects the fact that out-of-phase modes are intrinsically resilient against homogeneous noise fields.

Finally, we apply our one pulse sequences with variable offset electric fields and curvatures to minimize coherent displacement contributions, i.e., compensating (static) stray electric fields. 
In our experiments, residual stray fields displace the ions from the potential minimum to an equilibrium position given by the restoring force of the potential $\propto\omega^2$. If $\omega$ is switched, the instantaneous equilibrium position is shifted and, thus, the ions experience an effective force that excites (unwanted) coherent motion.
While coarse tuning is performed with single ions, fine tuning is done on the hybrid crystal with an axial ion separation of $\approx\SI{5.5}{\micro\meter}$.
From corresponding results, we deduce residual, homogeneous stray fields of \SI{1}{\volt/\meter} along the mid frequency mode.
Stray fields acting differentially on each ion, i.e., stray curvatures, are smaller than $5\cdot10^{-8}\,\SI{}{\volt/\micro\meter^{2}}$.

\clearpage
\begin{figure}\centering
	\includegraphics{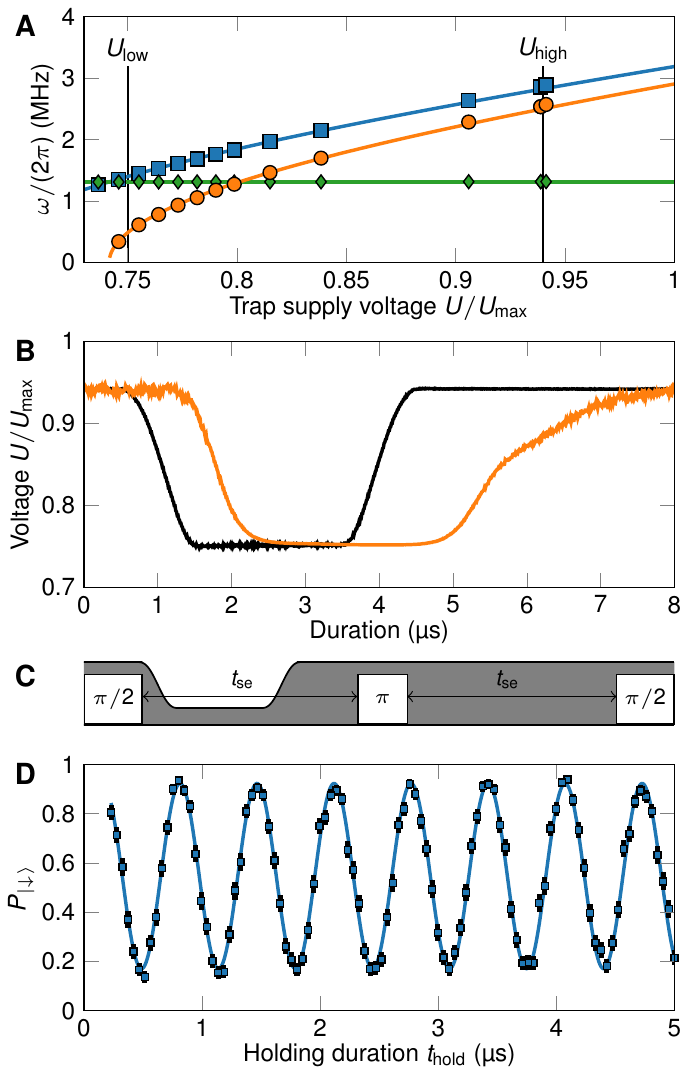}
	\caption{
		\textbf{Calibration of mode frequency tunability.}
		(\textbf{A})~
		Mode frequencies as a function of the rf voltage $U$, data points probed with a single ion and sem are smaller than symbols. 
		We detect the weaker radial mode frequency $\omega_\text{mf}$ (blue squares) and the axial frequency $\omega_\text{lf}$ (green diamonds).
		In addition we illustrate the estimated corresponding rocking mode frequency $\omega_\text{a}$ (orange circles) in the hybrid crystal.
		(\textbf{B})~
		Voltage amplitude $U$, estimated from the control voltage waveform $U_\text{c}(t)$
		(black) in comparison to the rf pickup $U_\text{p}$ (orange) for a typical pulse waveform with limits $U_\text{high}$ and $U_\text{low}$. The shape of $U_\text{p}$ is distorted by the limited bandwidth of components in the rf generator setup.
		(\textbf{C})~
		Motional spin-echo pulse sequence (not to scale) sensitive to relative mode frequency differences of both arms, each of length $t_\text{se}\propto t_\text{hold}$. We use this sequence to further benchmark the quality of our waveforms.
		(\textbf{D})~
		Results of the spin-echo sequence for variable $t_\text{hold}$, probing the relevant radial mode with a single ion. A model fit (solid line) yields a frequency difference of $2\pi\times$\SI{1.53(3)}{\mega\hertz}.
		\label{figS1}}
\end{figure}

\clearpage
\section{Motional state reconstruction}

The analysis for motional state reconstruction consists of two crucial steps: 
(I) Using well-established techniques to determine population distributions (data points in Fig. 3B, C and Fig. 4D) and (II) to yield estimates for squeezing and displacement excitations.
For the first step, we employ red or blue sideband couplings to the relevant rocking mode in individual sequences for variable durations $t_\text{sb}$, and record $P_{|\downarrow\rangle}$. 
Figure~S2 depicts the experimental data for the states depicted in Figs.~3 and 4 where each data point is the result of 200 repetitions.
We reconstruct the motional state distribution $P_n$ (population of each number state, data points in Figs.~3B-C and 4D) from combined model fits to the probability $P_{|\downarrow\rangle}(t)$ (solid lines), see~\cite{Leibfried2003}.
Note, sideband coupling (Rabi) rates depend on $\ket{n}$ and motional state distributions lead to characteristic evolutions of $P_{|\downarrow\rangle}$, and we choose to consider up to $n = 8$ in the reconstruction.
Independent from any further considerations, we note that, in particular, data points shown in Fig. 4D exhibit direct evidence for substantial squeezing and negligible displacement and thermal excitations. 
Only even number states are significantly occupied (particle pairs), while populations are in agreement with zero for odd states (unwanted excitations).

In the second step of our analysis, we estimate contributions of squeezing $r$ and displacement $\alpha$ (due to residual stray electric fields), while heating effects are measured to be small in our case (see above) and neglected in the following. 
To this end we define creation (annihilation) operators $\hat{a}^\dagger$ ($\hat{a}$) and combine a squeezing operator $\hat{S}(\xi) = \exp\left(\frac12\left(\xi^*\hat{a}^2-\xi \hat{a}^{\dagger 2}\right)\right)$, with $\xi=r e^{i\theta}$, and a displacement operator $\hat{D}(\alpha) = \exp\left(\alpha\hat{a}^\dagger-\alpha^*\hat{a}\right)$ to create a squeezed coherent state $\rho_\text{fin}$ by 
\begin{align}
	\rho_\text{fin} = \hat{S}(r)\hat{D}(\alpha) \rho_\text{ini} \hat{D}^\dagger(\alpha)\hat{S}^\dagger(r)\,.
\end{align}
Here, $\rho_\text{ini}(\bar{n}_\text{th})$ denotes the initial (thermal) density matrix. 
The diagonal elements of $\rho_\text{fin}$ represent the elements $P_n^\text{par}(r,\alpha,\bar{n}_\text{th})$, which we fit to the measured $P_n$ to estimate the final experimental state.
In particular, we allow for a differential phase between squeezing and displacement to reduce systematic deviations.
From $\rho_\text{fin}$ the amplitudes $r$ and $|\alpha|$ can be extracted similarly to the numerically simulated states (see below).
We note that it is also possible to reconstruct the full density matrix, i.e., Wigner function, experimentally~\cite{Leibfried1996}, from which, in turn, squeezing and coherent displacements could be distinguished directly.

\clearpage
\begin{figure}\centering
	\includegraphics{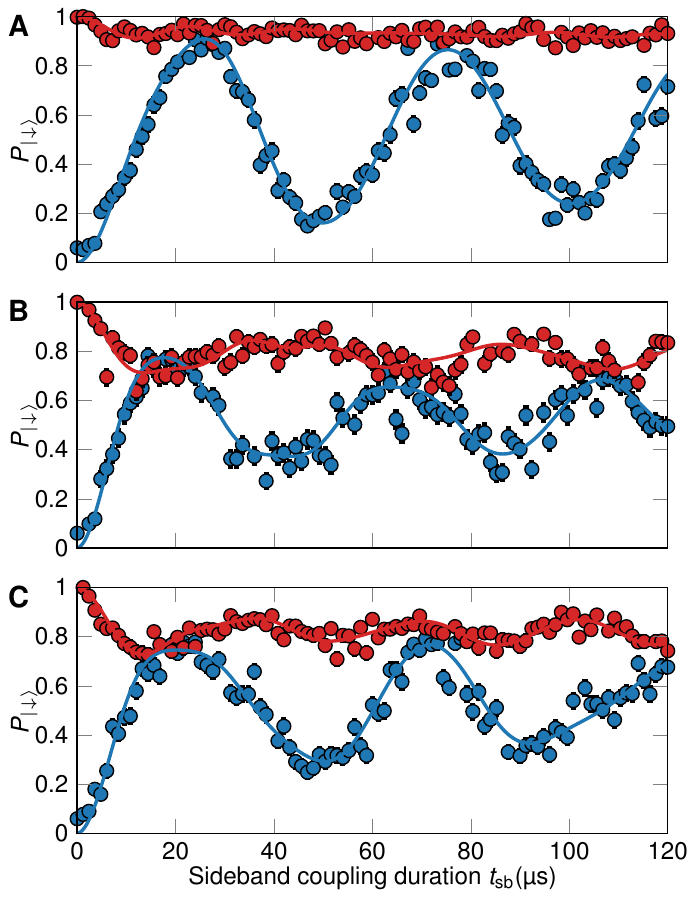}
	\caption{
		\textbf{Spin signals for reconstruction of the rocking mode population distribution.}
		We map the rocking mode state of the hybrid crystal onto the spin state of the $^{25}$Mg$^+$ via dedicated red (red data points) or blue (blue data points) sideband transitions. 
		Each data point corresponds to 200 repetitions of our experimental sequences (described in the main text), error bars depict the sem. 
		Solid lines show model fits yielding corresponding $P_n$:
		(\textbf{A}) Initial state and (\textbf{B}) excited motional state after the one-pulse sequence, cf. Figs. 3B and C.
		(\textbf{C})~
		Final motional state after purification via the echo sequence with $t_\text{free}=\SI{30.2}{\micro\second}$, cf. Fig. 4C.	
		\label{figS10}}
\end{figure}


\clearpage
\section{Numerical simulations}

In the following, we perform numerical simulations to gain an additional perspective on the underlying systematic effects in the relevant dynamics, in particular, the effect of finite ramp durations and the accuracy of $P_n^\text{par}(r,\alpha,\bar{n}_\text{th})$.
We consider a single harmonic oscillator  with time-dependent eigenfrequency $\omega(t)$ and recast the corresponding Hamiltonian~\cite{Silveri2015}:
\begin{align}
	\hat{H}(t) &= \hbar \omega(t)\left(\hat{a}_\omega^\dagger(t)\hat{a}_\omega(t)+\frac{1}{2}\right)\\
	&= \hbar \omega(t)\left(\hat{a}^\dagger\hat{a}+\frac{1}{2}\right) - \frac{i\hbar}{4}\frac{d \ln{\left(\omega(t)\right)}}{dt}\left[\hat{a}^2-\left(\hat{a}^\dagger\right)^2\right]
	\,.
\end{align}
Here, $\hat{a}_\omega^\dagger(t)$ and $\hat{a}_\omega(t)$ denote time-dependent creation and annihilation operators, that are transformed into time-independent basis operators $\hat{a}^\dagger$ and $\hat{a}$ defined for the initial mode frequency $\omega_0 = \omega(0)$.
Further, we include the effect of stray electric fields by an additional term:
\begin{align}
	\hat{H}_F(t) = x_0 F_0 (t)\,(\hat{a} + \hat{a}^\dagger)
	\,,
\end{align}
where $F_0(t)$ denotes an effective stray force amplitude and $x_0=\sqrt{\hbar/\left(2m\omega_0\right)}$ is the width of the ground state wave function of the oscillator with mass $m$. 
In addition, we define position $\hat{x} = \sqrt{\hbar/\left(2 m \omega_0\right)}\,\left(\hat{a}^\dagger+\hat{a}\right)$ and momentum $\hat{p} = i \sqrt{\hbar m \omega_0/2}\,\left(\hat{a}^\dagger-\hat{a}\right)$ operators, corresponding expectation values $x(t)=\langle \hat{x}\rangle$ and $p(t)=\langle \hat{p} \rangle$, and variances $\left(\Delta x\right)^2(t)=\langle\hat{x}^2\rangle-\langle \hat{x} \rangle^2$ and $\left(\Delta p\right)^2(t)=\langle\hat{p}^2\rangle-\langle \hat{p} \rangle^2$.
The squeezing and displacement amplitudes are then given by $r(t)=\frac12\arccosh\left[\left(\Delta x\right)^2(t)+\left(\Delta p\right)^2(t)\right]$ and $|\alpha(t)|=\sqrt{x^2(t)+p^2(t)}$, respectively.
We numerically simulate the total Hamiltonian $\hat{H} + \hat{H}_F$ using \texttt{QUTIP}~\cite{Johansson2013} and obtain unitary time-evolutions of the density matrix, from which we can calculate all of the defined expectation values, neglecting any decoherence effects.

First, we assess the effect of variable ramp durations and depict final $r$ for the single ramp, with fixed stray field amplitude, from low frequency $\omega_\text{low} = \omega_\text{a}(U_\text{low})$ to high frequency $\omega_\text{high} = \omega_\text{a}(U_\text{high})$ in Fig.~S3A. 
For shorter $t_\text{ramp}$, squeezing approaches the ultimate value, corresponding to an instantaneous switching, given by $r_\text{max}=\frac{1}{2}\ln\left(\frac{\omega_\text{high}}{\omega_\text{low}}\right) \approx 0.8 $. 
Assuming a variation of \SI{10}{\percent} of the ramp duration around \SI{1}{\micro\second}, the squeezing amplitude varies by less than \SI{10}{\percent}. 
In comparison, we illustrate variable contributions of final $|\alpha|$ as a function of $t_\text{ramp}$, for fixed stray field amplitudes, in Fig.~S3B. 

Second, we study the accuracy of $P_n^\text{par}(r,\alpha,\bar{n}_\text{th})$ for the case of our two-pulse sequence. 
To this end, we insert corresponding experimental parameter settings, determined by the above described calibration measurements, into the numerical simulations.
In Figure~S4, we show final numerical $r$ and $|\alpha|$ for variable changes of the free evolution duration $\Delta t_\text{free}$, including a fixed stray field (amplitude $\SI{11}{\milli\volt/\meter}$).
We fit numerical $P_n(\Delta t_\text{free})$ with $P_n^\text{par}(r,\alpha,\bar{n}_\text{th})$ to extract final $r$ (dashed orange line) and $|\alpha|$ (dashed gray line) and compare it to exact numerical results (solid lines).
In Figure~S4B we plot corresponding residuals and find overall agreement with residual deviations on the order of a few percent.

\clearpage
\begin{figure}\centering
	\includegraphics{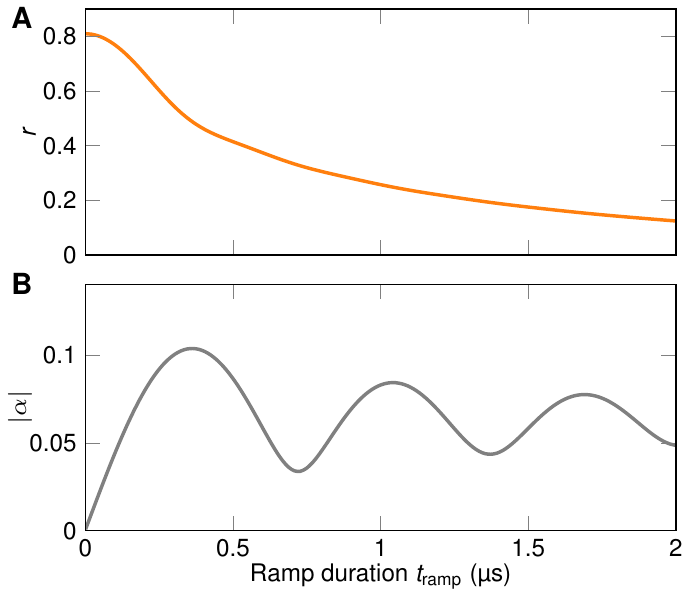}
	\caption{
		\textbf{Illustration of the effects of finite ramp durations.}
		(\textbf{A})~
		For ever shorter $t_\text{ramp}$ the squeezing amplitude approaches the ultimate value of an instantaneous switching $r_\text{max}\approx 0.8$.
		(\textbf{B})~
		The coherent displacement due to a residual stray field (amplitude \SI{11}{\milli\volt/\meter}) is modulated as a function of $t_\text{ramp}$, illustrating the evolution of the WKB phase during the ramps for finite $t_\text{ramp}$.
		\label{figS0}}
\end{figure}

\clearpage
\begin{figure}\centering
	\includegraphics{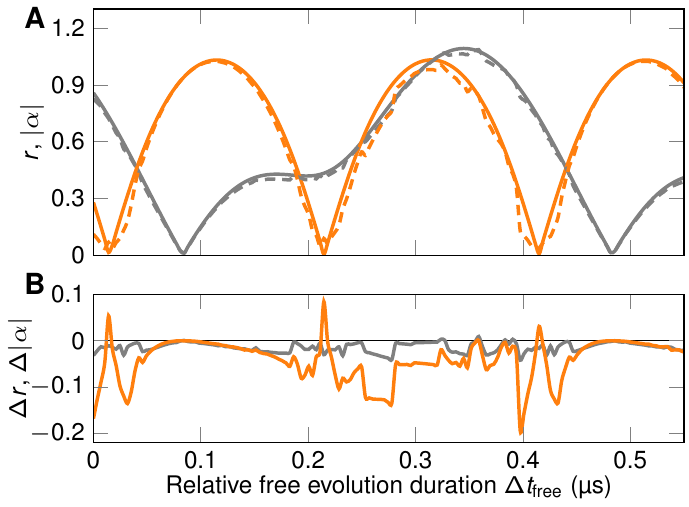}
	\caption{
		\textbf{Systematics of values extracted with our parametrized phonon number distribution.}
		(\textbf{A})~
		We depict simulated values for $r$ (orange solid line) and $|\alpha|$ (gray solid line) for the two-pulse sequence as a function of a relative duration $\Delta t_\text{free}$.
		Dashed lines depict the values obtained by fitting $P_n^\text{par}(r,\alpha,\bar{n}_\text{th})$ to the simulated $P_n$.
		(\textbf{B})~
		Residuals between fit parameters and simulated values for squeezing (orange) and coherent displacement (gray).
		\label{figS12}}
\end{figure}

\clearpage
\section{Spatial entanglement of two ions}

We present a summarizing view on the spatial entanglement in our bipartite system, ion A and B, that builds on original work of\,\cite{Retzker2005, Fey2018} and consider the specific condition in our experiment. 
Ions A and B are confined in three spatial degrees of freedom, interact individually with the common storage field, and with each other due to their mutual Coulomb interaction.
We focus on the description of one spatial degree of freedom, e.g., the weaker radial direction, and define individual, single ion oscillator states $\ket{n}_\text{A, B}$ with corresponding (uncoupled) eigenfrequencies $\omega_\text{A} = \omega_\text{B}$.
Due to the ion-ion interaction, both harmonic oscillators are coupled with rate $\omega_\text{coupl}$.
We distinguish short timescales $t \ll 1/\omega_\text{coupl}$ from long timescales $t \gg 1/\omega_\text{coupl}$, where the motion is typically described by a common in-phase and out-of-phase motion with eigenfrequencies $\omega_\text{+, -}$ and corresponding number states $\ket{n}_\text{+, -}$, respectively.
Further, we assume an initial, vacuum state~\cite{Retzker2005}:
\begin{align}
	\ket{\text{ini}} &= \ket{0}_\text{+}\ket{0}_\text{-} \\
	&= \sqrt{1-e^{-2\beta}} \sum_n e^{-\beta\,n} \ket{n}_\text{A}\ket{n}_\text{B},
	\label{eqn:vac_state}
\end{align}
with $e^{-\beta} = \sqrt{(\kappa - 1/2)/(\kappa + 1/2)}$ and $\kappa = 1/4\left( \sqrt{\omega_{+}/\omega_{-}}+\sqrt{\omega_{-}/\omega_{+}}\right)$. 
In the basis of $\ket{n}_\text{A, B}$, this corresponds to an entangled, squeezed state and, for our experimental parameters $\kappa \approx 0.5008$.
In our experiments, we increase the spatial entanglement by the coherent generation of squeezing in the out-of-phase mode, while squeezing of the in-phase mode is, estimated from numerical simulations, negligible.
We can separate the residual (classical) coherent displacement from the non-classical dynamics and, for illustration, we assume pure motional squeezing $\hat{S}(\xi)$ to estimate our final state:
\begin{align}
	\ket{\text{fin}} &= \ket{0}_\text{+}\ket{0}_\text{-} + \frac{\xi}{\sqrt{2}} \ket{0}_\text{+}\ket{2}_\text{-} + \mathcal{O}(r^2)\\
	&= \ket{0}_\text{A}\ket{0}_\text{B} - \frac{\xi}{2}  \ket{1}_\text{A}\ket{1}_\text{B} - \frac{\xi}{\sqrt{8}} \left(\ket{0}_\text{A} \ket{2}_\text{B}+\ket{2}_\text{A} \ket{0}_\text{B}\right)\  + \mathcal{O}(r^2).
	\label{eqn:entangled_state}
\end{align}
Note, the second term is related to our state illustration $\ket{{}^\bullet\,{}_\bullet}+\ket{{}_\bullet\,{}^\bullet}$ given in the main text.

To quantify the amount of entanglement, we, further, consider that the ions are initially prepared in a thermal state described by average phonon numbers $\bar{n}_\text{+, -}$ and states remain Gaussian throughout our experimental sequences. 
The entanglement of formation of the ions is then given by~\cite{Adesso2007,Serafini2004,Bruschi2013,Fey2018}:
\begin{equation}
	E_F(\chi)=\frac{(1/2+\chi)^2}{2\chi} \ln\left( \frac{(1/2+\chi)^2}{2\chi}\right)-\frac{(1/2-\chi)^2}{2\chi} \ln\left( \frac{(1/2-\chi)^2}{2\chi}\right),
\end{equation}
where $\chi$ is related to the so-called symplectic eigenvalues:
\begin{align}
	\lambda_1&=e^{-r} \frac{\sqrt{1+2\bar{n}_\text{-}} \sqrt{1+2\bar{n}_\text{+}} \sqrt{\omega_\text{+}}}{2 \sqrt{\omega_\text{-}}}\label{eqn:symplectic_eigenvalue1}\\
	\lambda_2&=e^{r} \frac{\sqrt{1+2\bar{n}_\text{-}} \sqrt{1+2\bar{n}_\text{+}} \sqrt{\omega_\text{-}}}{2 \sqrt{\omega_\text{+}}}
	\label{eqn:symplectic_eigenvalue2}
\end{align}
via $\chi=\min(\lambda_1,\lambda_2)$.
The symplectic eigenvalues and, therefore, $E_F$ are functions of the applied squeezing $r$, i.e., $E_F (r=0) \approx 10^{-5}$ for our initial state and $E_F (r=0.83) \approx 0.41$ for our final state in Fig.~4D.
Note, Equations~\eqref{eqn:symplectic_eigenvalue1} and \eqref{eqn:symplectic_eigenvalue2} are a generalization of results obtained in~\cite{Fey2018}, which take the impact of finite coupling strength $\omega_+/\omega_-$ into account. 
For pure states and in the absence of squeezing, i.e, $\bar{n}_- = \bar{n}_+ = r = 0$, the entanglement of formation Eq.~(S21) reduces to the von Neumann entanglement as stated in~\cite{Retzker2005}.  


\begin{thebibliography}{10}
	
	\bibitem{Lamb1947}
	W.~E. Lamb, R.~C. Retherford, {\it Physical Review\/} {\bf 72}, 241 (1947).
	
	\bibitem{Casimir1948}
	H.~B.~G. Casimir, {\it Proc. Kon. Ned. Akad. Wetensch.\/} {\bf 51}, 793 (1948).
	
	\bibitem{Einstein1917}
	A.~Einstein, {\it Physikalische Zeitschrift\/} {\bf 18}, 121 (1917).
	
	\bibitem{Dirac1927}
	P.~A.~M. Dirac, {\it Proceedings of the Royal Society A\/} {\bf 114}, 243
	(1927).
	
	\bibitem{Schroedinger1939}
	E.~Schr{\"{o}}dinger, {\it Physica\/} {\bf 6}, 899 (1939).
	
	\bibitem{Parker1969}
	L.~Parker, {\it Physical Review\/} {\bf 183}, 1057 (1969).
	
	\bibitem{Sauter1931}
	F.~Sauter, {\it Zeitschrift f{\"{u}}r Physik\/} {\bf 69}, 742 (1931).
	
	\bibitem{Hawking1974}
	S.~W. Hawking, {\it Nature\/} {\bf 248}, 30 (1974).
	
	\bibitem{Moore1970}
	G.~T. Moore, {\it Journal of Mathematical Physics\/} {\bf 11}, 2679 (1970).
	
	\bibitem{Mukhanov1992}
	V.~Mukhanov, H.~Feldman, R.~Brandenberger, {\it Physics Reports\/} {\bf 215},
	203 (1992).
	
	\bibitem{Belgiorno2010}
	F.~Belgiorno, {\it et~al.\/}, {\it Physical Review Letters\/} {\bf 105}, 203901
	(2010).
	
	\bibitem{Lahav2010}
	O.~Lahav, {\it et~al.\/}, {\it Physical Review Letters\/} {\bf 105}, 240401
	(2010).
	
	\bibitem{Wilson2011}
	C.~M. Wilson, {\it et~al.\/}, {\it Nature\/} {\bf 479}, 376 (2011).
	
	\bibitem{Weinfurtner2011}
	S.~Weinfurtner, E.~W. Tedford, M.~C.~J. Penrice, W.~G. Unruh, G.~A. Lawrence,
	{\it Physical Review Letters\/} {\bf 106}, 021302 (2011).
	
	\bibitem{Jaskula2012}
	J.~C. Jaskula, {\it et~al.\/}, {\it Physical Review Letters\/} {\bf 109},
	220401 (2012).
	
	\bibitem{Laehteenmaeki2013}
	P.~L{\"{a}}hteenm{\"{a}}ki, G.~S. Paraoanu, J.~Hassel, P.~J. Hakonen, {\it
		Proceedings of the National Academy of Sciences\/} {\bf 110}, 4234 (2013).
	
	\bibitem{Steinhauer2016}
	J.~Steinhauer, {\it Nature Physics\/} {\bf 12}, 959 (2016).
	
	\bibitem{Euve2016}
	L.~P. Euv{\'{e}}, F.~Michel, R.~Parentani, T.~G. Philbin, G.~Rousseaux, {\it
		Physical Review Letters\/} {\bf 117}, 121301 (2016).
	
	\bibitem{Eckel2018}
	S.~Eckel, A.~Kumar, T.~Jacobson, I.~B. Spielman, G.~K. Campbell, {\it Physical
		Review X\/} {\bf 8}, 021021 (2018).
	
	\bibitem{Leibfried2003}
	D.~Leibfried, R.~Blatt, C.~Monroe, D.~Wineland, {\it Review of Modern
		Physics\/} {\bf 75}, 281 (2003).
	
	\bibitem{Wineland2013}
	D.~J. Wineland, {\it Reviews of Modern Physics\/} {\bf 85}, 1103 (2013).
	
	\bibitem{Alsing2005}
	P.~M. Alsing, J.~P. Dowling, G.~J. Milburn, {\it Physical Review Letters\/}
	{\bf 94}, 220401 (2005).
	
	\bibitem{Schuetzhold2007}
	R.~Sch{\"{u}}tzhold, {\it et~al.\/}, {\it Physical Review Letters\/} {\bf 99},
	201301 (2007).
	
	\bibitem{Fey2018}
	C.~Fey, T.~Schaetz, R.~Sch{\"{u}}tzhold, {\it Physical Review A\/} {\bf 98},
	33407 (2018).
	
	\bibitem{Supp}
	{See the Supplemental Material.}
	
	\bibitem{Martin2012}
	E.~Mart{\'{i}}n-Mart{\'{i}}nez, N.~C. Menicucci, {\it Classical and Quantum
		Gravity\/} {\bf 29}, 224003 (2012).
	
	\bibitem{Clos2016}
	G.~Clos, D.~Porras, U.~Warring, T.~Schaetz, {\it Physical Review Letters\/}
	{\bf 117}, 170401 (2016).
	
	\bibitem{Serafini2004}
	A.~Serafini, F.~Illuminati, S.~D. Siena, {\it J. Phys. B: At. Mol. Opt.
		Phys.\/} {\bf 37}, L21 (2004).
	
	\bibitem{Retzker2005}
	A.~Retzker, J.~I. Cirac, B.~Reznik, {\it Physical Review Letters\/} {\bf 94},
	050504 (2005).
	
	\bibitem{Jost2009}
	J.~D. Jost, {\it et~al.\/}, {\it Nature\/} {\bf 459}, 683 (2009).
	
	\bibitem{Burd2018}
	S.~C. Burd, {\it et~al.\/}, {\it arXiv:1812.01812\/}  (2018).
	
	\bibitem{Cirac2012}
	J.~I. Cirac, P.~Zoller, {\it Nature Physics\/} {\bf 8}, 264 (2012).
	
	\bibitem{Ge2019}
	W.~Ge, {\it et~al.\/}, {\it Physical Review Letters\/} {\bf 122}, 030501
	(2019).
	
	\bibitem{Fluehmann2019}
	C.~Fl{\"{u}}hmann, {\it et~al.\/}, {\it Nature\/} {\bf 566}, 513 (2019).
	
	\bibitem{Cirac2000}
	J.~I. Cirac, P.~Zoller, {\it Nature\/} {\bf 404}, 579 (2000).
	
	\bibitem{Kielpinski2002}
	D.~Kielpinski, C.~Monroe, D.~J. Wineland, {\it Nature\/} {\bf 417}, 709 (2002).
	
	\bibitem{Birrell1982}
	N.~D. Birrell, P.~C.~W. Davies, {\it {Quantum fields in curved space}\/}
	(Cambridge University Press, Cambridge, 1982).
	
	\bibitem{Schutzhold2013}
	R.~Sch{\"{u}}tzhold, W.~G. Unruh, {\it Physical Review D\/} {\bf 88}, 124009
	(2013).
	
	\bibitem{Bowler2013}
	R.~Bowler, U.~Warring, J.~W. Britton, B.~C. Sawyer, J.~Amini, {\it Review of
		Scientific Instruments\/} {\bf 84}, 033108 (2013).
	
	\bibitem{Schmitz2009}
	H.~Schmitz, {\it et~al.\/}, {\it Applied Physics B\/} {\bf 95}, 195 (2009).
	
	\bibitem{Wittemer2017}
	M.~Wittemer, G.~Clos, H.-P. Breuer, U.~Warring, T.~Schaetz, {\it Physical
		Review A\/} {\bf 97}, 020102 (2017).
	
	\bibitem{Leibfried2002}
	D.~Leibfried, {\it et~al.\/}, {\it Physical Review Letters\/} {\bf 89}, 247901
	(2002).
	
	\bibitem{Leibfried1996}
	D.~Leibfried, {\it et~al.\/}, {\it Physical Review Letters\/} {\bf 77}, 4281
	(1996).
	
	\bibitem{Silveri2015}
	M.~P. Silveri, J.~A. Tuorila, E.~V. Thuneberg, G.~S. Paraoanu, {\it Reports on
		Progress in Physics\/} {\bf 80}, 056002 (2015).
	
	\bibitem{Johansson2013}
	J.~Johansson, P.~Nation, F.~Nori, {\it Computer Physics Communications\/} {\bf
		184}, 1234 (2013).
	
	\bibitem{Adesso2007}
	G.~Adesso, F.~Illuminati, {\it J. Phys. A: Math. Theor.\/} {\bf 40}, 7821
	(2007).
	
	\bibitem{Bruschi2013}
	D.~E. Bruschi, N.~Friis, I.~Fuentes, S.~Weinfurtner, {\it New Journal of
		Physics\/} {\bf 15}, 113016 (2013).
	
\end{thebibliography}
\end{document}